\RequirePackage[2020-02-02]{latexrelease}
\documentclass[aps,amsmath,amssymb,superscriptaddress,preprint,hyperref]{revtex4}
\usepackage{graphicx}
\usepackage{xcolor}

\def\gsim{\mathrel{\rlap{\lower4pt\hbox{\hskip1pt$\sim$}}
    \raise1pt\hbox{$>$}}}       %greater than or approx. symbol

\def\Im{\,{\rm Im}\, }

\def\phl{\vphantom{l}}
\def\a{\alpha}
\def\b{\beta}

\def\e{\epsilon}

\def\m{\mu}
\def\n{\nu}

\def\r{\rho}

\def\s{\sigma}

\def\et{\eta}

\def\D{\Delta} 

\newcommand{\gn}{G_{\rm N}}

\begin{document}
\title{Matching the Vilkovisky-DeWitt Effective Action of Quantum Gravity to String Theory}
\author{Xavier Calmet} \email{x.calmet@sussex.ac.uk}
\affiliation{Department of Physics and Astronomy,\\
    University of Sussex, Brighton, BN1 9QH, United Kingdom}	
    \author{Elias Kiritsis}\email{kiritsis@physics.uoc.gr}\affiliation{Universit\' e Paris Cit\' e, CNRS, Astroparticule et Cosmologie, F-75013 Paris, France.
}\affiliation{Crete Center for Theoretical Physics, Institute for Theoretical and Computational Physics, Department of Physics
University of Crete, Heraklion, Greece.}
\affiliation{Arnold Sommerfeld Center for Theoretical Physics,\\
	Ludwig-Maximilians-Universit\"at M\"unchen, 80333 M\"unchen, Germany.}
\author{Folkert Kuipers} \email{f.kuipers@physik.uni-muenchen.de}
\affiliation{Arnold Sommerfeld Center for Theoretical Physics,\\
    Ludwig-Maximilians-Universit\"at M\"unchen, 80333 M\"unchen, Germany.}
\author{Dieter L\"ust}\email{luest@mppmu.mpg.de}
\affiliation{Arnold Sommerfeld Center for Theoretical Physics,\\
    Ludwig-Maximilians-Universit\"at M\"unchen, 80333 M\"unchen, Germany.}
\affiliation{Max-Planck-Institut f\"ur Physik (Werner-Heisenberg-Institut),
	Boltzmannstr. 8, 85748 Garching, Germany.}

\preprint{LMU-ASC 12/24}

\preprint{MPP-2024-172}
\preprint{CCTP-2024-13}

\preprint{ITCP-2024/13}

\begin{abstract}
In this work, we discuss the matching of the Vilkovisky-DeWitt  effective action of quantum gravity to an example of an ultra-violet complete theory of quantum gravity. We show how this matching enables a calculation of the local Wilson coefficients of the effective action.
We provide several examples in string theory. Moreover, we discuss the matching conditions within the context of the swampland program.
\end{abstract}

\maketitle

The aim of the paper is to discuss the matching of the Vilkovisky-DeWitt effective action of quantum gravity to an ultra-violet complete theory of quantum gravity, namely string theory. Besides explicitly demonstrating how to match this effective field theory to string theory, our work enables tests of string theory in low energy experiments which are sensitive to gravity by connecting an effective action obtained from string theory to low energy physical observables. Finally, we comment on the case of large species number and describe how to evaluate the Wilson coefficients of the effective action in that situation.

The Vilkovisky-DeWitt effective action of quantum gravity \cite{Barvinsky:1983vpp,Barvinsky:1985an,Barvinsky:1987uw,Barvinsky:1990up,Buchbinder:1992rb} is obtained by integrating out the quantum fluctuations of the graviton and potentially other massless fields. As this integration is performed in such a way that the resulting effective action is independent of the gauge fixing and the parameterization of the fields \cite{FN1}, this effective action is known as the unique effective action of quantum gravity \cite{Barvinsky:1983vpp,Vilkovisky:1984st}. As such the action enables to do calculations in quantum gravity at energies well below the Planck scale (or equivalently for weak curvatures) in a consistent and systematic manner without assuming a specific ultra-violet completion. 

While this approach enables model independent predictions in quantum gravity when the observables under consideration only depend on the  non-local part of the action, cf. e.g. Refs.~\cite{Donoghue:1993eb,Donoghue:1994dn,Calmet:2021lny,Calmet:2021stu,Calmet:2018uub,Calmet:2018rkj,Calmet:2016sba}, it does not solve the non-renormalizability of quantum gravity, the Wilson coefficients of the local part of the action cannot be calculated from first principles without assuming an ultra-violet theory of quantum gravity. In this paper, we show how to match this effective action to string theory in the weak string coupling limit (see e.g., \cite{Gregori:1997hi,Alvarez-Gaume:2015rwa}). This matching enables a calculation of the Wilson coefficients of the local part of the unique effective action. We show by considering several examples in string theory that these coefficients depend on the details of the compactification of extra-dimensions. After discussing the matching of the effective actions for these explicit examples, we discuss how the local Wilson coefficients are generically determined by UV cut-off scale, where gravity becomes strongly coupled, as determined by the species scale \cite{Dvali:2007hz,Dvali:2007wp,Dvali:2009ks,Dvali:2010vm,Dvali:2012uq,Calmet:2008tn,Atkins:2010eq,Han:2004wt,Calmet:2014gya}.

The quantum corrections to classical solutions of general relativity are reliably calculable using quantum corrected field equations obtained from the variation of the unique effective action of quantum gravity, as long as curvature invariants remain weak.  At second order in curvature, the unique effective action is given by \cite{Barvinsky:1983vpp,Barvinsky:1985an,Barvinsky:1987uw,Barvinsky:1990up,Buchbinder:1992rb,Calmet:2018elv}
\begin{equation}
    \Gamma_{\rm QG} = \Gamma_{\rm L} + \Gamma_{\rm NL} +\Gamma_{\rm matter}
\end{equation}
with a local part
\begin{align}
    \Gamma_{\rm L}
    &=
    \int d^4x \, \sqrt{|g|} \left[ \frac{M_P^2 }{2}
    \big(R - 2\Lambda\big)
    + c_1(\mu) \, R^2
    + c_2(\mu) \, R_{\mu\nu} R^{\mu\nu}
    \right.\nonumber\\
    &\qquad \qquad \qquad \qquad
    + c_3(\mu) \, R_{\mu\nu\rho\sigma} R^{\mu\nu\rho\sigma}
    + c_4(\mu) \, \Box R
    + \mathcal{O}(M_P^{-2}) \Big]
\end{align}
and a non-local part
\begin{align}
    \Gamma_{\rm NL} &=
    - \int d^4x \, \sqrt{|g|} \left[
    b_1 \, R \ln \left(\frac{\Box}{\mu^2} \right) R
    + b_2 \, R_{\mu\nu} \ln \left( \frac{\Box}{\mu^2} \right) R^{\mu\nu}
    \right.\nonumber\\
    &\qquad \qquad \qquad \qquad \left.
    + b_3 \, R_{\mu\nu\rho\sigma} \ln \left( \frac{\Box}{\mu^2} \right) R^{\mu\nu\rho\sigma}
    + \mathcal{O}(M_P^{-2}) \right],
\end{align}
and where $\Gamma_{\rm matter}$ is the matter part of the effective action.

Here, $M_P=\sqrt{\hbar \, c/(8 \,\pi \, \gn)}= 2.4\times10^{18}\,{\rm GeV}$ denotes the reduced Planck mass, $\Lambda$ the cosmological constant and $c_i$ and $b_i$ are Wilson coefficients. Furthermore, the parameter $\mu$ is a renormalization scale. As always in quantum field theory, physical observables do not depend on this renormalization scale, even though its dependence is explicit in the action.
\par

We emphasize that this effective action where all quantum effects in a fixed external background have been calculated by integrating out modes at all scales, and the result has been 
Legendre-transformed. This is unlike the Wilsonian effective action that comes with a cutoff, and where all degrees of freedom above the cutoff have been integrated out while those below the cutoff not. As such, the gravitational sector of the unique effective action is purely classical. 

In the remainder, we will set the cosmological constant to zero \cite{FN2}. In addition, we ignore the boundary term associated with $c_4$, as it does not contribute to the field equations. Furthermore, the non-local Wilson coefficients are related to infrared divergences, thus do not depend on the ultra-violet completion of the theory. Hence, they can be calculated, which has been done \cite{FN3}. The results are given by
\begin{align}
    b_1 &= \frac{5(6\xi-1)^2 \, N_S -5 \, N_F - 50 \, N_V + 250 \, N_G }{11520 \pi^2} \, , \label{b1}\\
    b_2 &= \frac{-2 \, N_S + 8 \, N_F + 176 \, N_V - 244 \, N_G }{11520 \pi^2} \, , \label{b2}\\
    b_3 &= \frac{2 \, N_S + 7 \, N_F - 26 \, N_V + 424 \, N_G }{11520 \pi^2} \, , \label{b3}
\end{align}
where $N_S$, $N_F$, $N_V$ and $N_G$ denote respectively the number of light scalars, fermions, vectors and gravitons in the theory, and $\xi$ denotes the non-minimal coupling of the scalar fields to scalar curvature.

The coefficients of the local part of the effective action depend on the renormalization scale as
\begin{align}\label{runc1}
    c_1(\mu) &= c_1(\mu_\ast) - b_1 \, \ln\left( \frac{\mu^2}{\mu_\ast^2}\right), \\ \label{runc2}
    c_2(\mu) &= c_2(\mu_\ast) - b_2 \, \ln\left( \frac{\mu^2}{\mu_\ast^2}\right), \\ \label{runc3}
    c_3(\mu) &= c_3(\mu_\ast) - b_3 \, \ln\left( \frac{\mu^2}{\mu_\ast^2}\right),
\end{align}
where $\mu_\ast$ is the scale at which effective action is matched to the ultra-violet complete theory of quantum gravity \cite{Buchbinder:1992rb,Lombardo:1996gp}. As the local coefficients $c_i(\mu_\ast)$ are related to the ultra-violet divergences of the theory, they can only be calculated within a specific ultra-violet completion of gravity. However, there are some bounds on these parameters. These bounds follow from the fact that the quadratic terms in the effective action introduce a massive scalar and a massive spin-2 mode into the theory \cite{Stelle:1976gc,Stelle:1977ry}, and the masses of these modes depend on the Wilson coefficients, cf. e.g. Ref. \cite{Calmet:2018qwg}. Hence, constraints on these masses introduce constraints on particular linear combinations of the coefficients.

At low energies, the masses of these modes are constrained by torsion balance experiments such as the E\"ot-Wash \cite{Hoyle:2004cw}, which yields a constraint of the form
\begin{align}
    |3\, c_1(0) + c_2(0) + c_3(0)| &\lesssim 10^{61} \, ,\nonumber\\
    |c_2(0) + 4\, c_3(0)| &\lesssim 10^{61} \, .
\end{align}
From an  effective field theory perspective, assuming that the scale of quantum gravity has been correctly identified, the local Wilson coefficients at the scale of quantum gravity $M_{\rm QG}$ are expected to be of the order
\begin{equation}
	c_i(M_{\rm QG})=\mathcal{O}(1) \, .
\end{equation}
However, as emphasized before the local Wilson coefficients cannot be calculated from first principles within the effective action framework and we need to match the unique effective action to an ultra-violet complete theory of quantum gravity to obtain reliable predictions for these local Wilson coefficients. 

Another important fact to keep in mind is that in models with a large number of species, the scale of quantum gravity $M_{QG}$ can be well below the reduced Planck mass. In the presence of $N$ light particles, the UV cut-off scale, where gravity becomes strongly coupled, is determined by the species scale $\Lambda_s$ (see \cite{Dvali:2007hz,Dvali:2007wp,Dvali:2009ks,Dvali:2010vm,Dvali:2012uq} where the term species scale is first introduced, see also \cite{Calmet:2008tn,Atkins:2010eq,Han:2004wt,Calmet:2014gya} where the dependence of $M_{QG}$ on $N$ is discussed  within the context of effective field theories, or moduli dependent species scale \cite{vandeHeisteeg:2022btw,Cribiori:2022nke,vandeHeisteeg:2023ubh,Cribiori:2023ffn,Cribiori:2023sch,vandeHeisteeg:2023dlw,Calderon-Infante:2023uhz,Castellano:2023aum,Basile:2023blg,Basile:2024dqq}). Using both perturbative and non-perturbative arguments one can show that this scale is given by
\begin{equation}
	\Lambda_s \leq \frac{M_P}{\sqrt {N}}\, .
\end{equation}
\par

From the effective field theory perspective, the species scale is obtained from the following argument. Studying tree level graviton scattering with matter fields as in and out states, one can show that the amplitudes grow with $E^2 N/M_P^2$ where $N$ is the number of fields in the matter sector of the model \cite{Han:2004wt,Atkins:2010eq}. Perturbative unitarity is thus violated at a scale $M_P/\sqrt{N}$. It can be shown that this is actually a sign of strong dynamics, and in the case of gravity that the species scale thus corresponds to the actual scale of quantum gravity \cite{Calmet:2014gya}. This is related to the self-healing mechanism \cite{Aydemir:2012nz,Calmet:2013hia} which shows that by resumming loops on the graviton propagator line in the large $N$ limit leads to an amplitude which fulfils the optical theorem. From the effective action point of view, the species scale thus arises dynamically.

The values of the local Wilson coefficients depend on the number of light species that are integrated out. Indeed, Eqs. \eqref{b1}, \eqref{b2} and \eqref{b3} show that the non-local Wilson coefficients $b_i$ are linearly dependent on the number $N$ of fields in the model. It follows from Eqs. \eqref{runc1}, \eqref{runc2} and \eqref{runc3} that the local Wilson coefficients $c_i$ scale with $N \log \mu$. At a low energy scale, they are thus also linearly proportional to $N$.

After this lightning review of the unique effective action approach to quantum gravity, we can now consider its matching to an ultra-violet complete theory of quantum gravity. 
As a prototype for such a theory, we consider type IIA, B string theory compactified  on K3 $\times$ T$^2$ to 4 dimensions. As explained in \cite{Gregori:1997hi}, the two-derivative low-energy effective action for $N = 4$ theories is believed to be exact at tree level, but higher-derivative terms can receive perturbative  one-loop corrections. We are particularly interested in the terms at second order in curvature \cite{Gregori:1997hi}:
\begin{eqnarray}
    {\cal I}_{\rm eff}
= \int d^4 x \sqrt{-g} & \Big( &\! \! \! \!
\Delta_{\rm gr} (T,U) R_{\mu\nu\rho\sigma} R^{\mu\nu\rho\sigma} +
\Theta_{\rm gr} (T,U) \epsilon^{\mu\nu\rho\sigma}_{\phl} R_{\mu\nu\alpha\beta}^{\phl}
R_{\rho\sigma}^{\ \ \alpha\beta}
\cr
&+& \! \! \! \!
\Delta_{\rm as} (T,U) \nabla_\mu H_{\nu\r\s}\nabla^\mu H^{\nu\r\s} +
\Theta_{\rm as} (T,U) \e^{\m\n\r\s}_{\phl}
\nabla_\mu^{\phl}H_{\nu\a\b}^{\phl} \nabla_{\r}^{\phl}
H_\s^{\hphantom{\r}\a\b} \cr
&+&\! \! \! \!
\Delta_{\rm dil} (T,U) \nabla_\m \nabla_\n \Phi \nabla^\m \nabla^\n \Phi +
\Theta_{\rm dil-as}(T,U) \e^{\m\n\r\s} \nabla_\m \nabla_\a \Phi \nabla^\a
H^{\n\r\s}
\cr
&+&\! \! \! \!
\Theta_{\rm gr-as} (T,U) \e^{\m\n\r\s}_{\phl} R_{\m\n\a\b}^{\phl}
\nabla_{\r}^{\phl} H_\s^{\hphantom{\r}\a\b}\,
\Big)\, ,
\label{eac}
\end{eqnarray}
where $\Delta$ denotes the couplings of the CP-even terms and $\Theta$ the couplings of the CP-odd terms, $H$ is the field strength of the $B$-field, $\Phi$ is the dilaton, $T$ denotes the K\"ahler moduli and $U$ the complex structure moduli.
After applying the Gauss-Bonnet identity, the CP-even purely gravitational part of the action is given by
\begin{eqnarray}\label{StringEFT}
    S_{ST}=\int_{M_4} d^4x \sqrt{-g} \left ( \frac{M_P^2}{2} R - \frac{1}{4 g_0^2} R^2+ \frac{1}{g_0^2} R_{\mu\nu}R^{\mu\nu} \right)
\label{n4}\end{eqnarray}
to second order in curvature expansion where
$g_0^2=\Delta_{\rm gr} (T,U)^{-1}$ and
\begin{eqnarray}
\mbox{type IIA:}\; \;  \D_{\rm gr} (T) =  -36 \log \left(
T_2 \left|
\et(T)\right|^4 \right)
 + {\rm const.}\, ,
\label{789}
\end{eqnarray}
\begin{eqnarray}
\mbox{type IIB:}\; \;  \D_{\rm gr} (U) = -36  \log \left(
U_2 \left|
\et(U)\right|^4 \right)
 + {\rm const.}\, .
\end{eqnarray}
Here $T$ and $U$ are the moduli of the two-torus and they can be calculated in terms of fundamental parameters of a specific compactification. They depend among other things on the masses of KK states and the lower dimensional Planck scale. We stress that the Wilson coefficients of the effective action derived from string theory are the moduli dependent number of particles $N$ in the definition of the species scale
\cite{vandeHeisteeg:2022btw,Cribiori:2022nke,vandeHeisteeg:2023ubh,Cribiori:2023ffn,Cribiori:2023sch,vandeHeisteeg:2023dlw,Calderon-Infante:2023uhz,Castellano:2023aum,Basile:2023blg,Basile:2024dqq}.

It is important to note that this action is obtained by integrating out all massive states in the string theory. Thus, it has a concrete cutoff that in generic situations is the string scale, $M_s$, and in case where compactifications volume and other moduli are larger, the mass of the lightest state that is moduli dependent. In the example of Eq. \eqref{n4} the UV cutoff of the Wilsonian effective action is generically ${M_s/ \Im T}$.

As another example, we could consider the effective action of string theory on a non-compact Calabi-Yau three-fold, $CY_3$, in the weak string coupling limit given by
\begin{equation}
    l_{s}^2 \rightarrow 0\,, \quad  g_s\rightarrow 0  \quad {\rm and} \quad \omega/l_s^2  <\infty \, ,
\end{equation}
where $g_s$ is the string coupling and $l_s$ the string length. Furthermore, $\omega$ is a parameter depending on the specifics of the Calabi-Yau three-fold, see below.
To second order in curvature, one has an action which has the same form as that of Eq.~\eqref{StringEFT}
with \cite{Alvarez-Gaume:2015rwa}
\begin{eqnarray}\label{defMP}
    M_P^2&=& -\frac{2 \zeta(3) \bar \chi}{3 (4 \pi)^7 l_s^2 g_s^2}\, , \\
    g_0^2&=& \frac{3 (4 \pi)^7 l_s^2}{4 \omega \zeta(3)}\, ,
\end{eqnarray}
where $\zeta$ is the Riemann zeta function. Moreover, $\omega$ and $\bar{\chi}$ are parameters depending on the specifics of the Calabi-Yau three-fold. For example, in the case of a compactification on a complex line bundle over $\mathbb{CP}^2$, they are given by \cite{Alvarez-Gaume:2015rwa}
\begin{align}
    \omega &= 384 \pi ^3 l^2 \, ,\\
    \bar{\chi} &= - 16384 \pi^3 \, ,
\end{align}
where $l$ is a free parameter that determines the minimal size of a four cycle diffeomorphic to $\mathbb{CP}^2$.

Remarkably, whether we consider compact or non-compact compactification manifolds, we end up with an effective action where the Wilson coefficients of the second order in curvature terms are related. Using the Gauss-Bonnet identity, $S_{ST}$ can be rewritten as \cite{FN4}
\begin{eqnarray}
    S_{ST}=\int_{M_4} d^4x \sqrt{-g} \left ( \frac{M_P^2}{2} R + \frac{1}{4 g_0^2} R^2 - \frac{1}{g_0^2} R_{\mu\nu}R^{\mu\nu} + \frac{1}{2g_0^2} R_{\mu\nu\alpha\beta}R^{\mu\nu\alpha\beta} \right).
\end{eqnarray}
We can thus read off the values of the $c_i$ at the string scale $M_S=1/l_s^2$ (note that the relation between the string scale and the Planck scale is given by Eq. (\ref{defMP})), we have:
\begin{eqnarray}
    c_1(M_S)&=& \frac{1}{4 g_0^2} \, ,\\
    c_2(M_S)&=& - \frac{1}{g_0^2} \, ,\\
    c_3(M_S)&=& \frac{1}{2g_0^2} \, .
\end{eqnarray}

The renormalization group equations of $c_1(\mu)$, $c_2(\mu)$  and $c_3(\mu)$ below the string scale are given by
\begin{align}
    c_1(\mu) &=  \frac{1}{4 g_0^2} - b_1 \, \ln\left( \frac{\mu^2}{M_S^2}\right), \\
    c_2(\mu) &= - \frac{1}{g_0^2} - b_2 \, \ln\left( \frac{\mu^2}{M_S^2}\right), \\
    c_3(\mu) &=  \frac{1}{2g_0^2} - b_3 \, \ln\left( \frac{\mu^2}{M_S^2}\right).
\end{align}
The Wilson coefficients of the local part of the action are thus calculable by matching the unique effective action to the ultra-violet complete theory of quantum gravity in the form of effective actions derived from different models of string theory.
In addition, the effective action allows to run down the values of the Wilson coefficients computed at the string scale to low energy scales. In principle, this allows to measure certain parameters of the string theory and its compactification in experiments performed at low energies. However, the constraints coming from current experiments are too weak to provide a meaningful restriction.

Now that we have discussed the matching of the unique effective field theory to concrete examples in string theory, we turn our attention to implications of the swampland program for the effective action. The expectations discussed above for the Wilson coefficients within the context of effective field theory are in full agreement with the value expected from swampland arguments. 
Indeed, according to the swampland program \cite{vandeHeisteeg:2022btw,Cribiori:2022nke,vandeHeisteeg:2023ubh,Cribiori:2023ffn,Cribiori:2023sch,vandeHeisteeg:2023dlw,Calderon-Infante:2023uhz,Castellano:2023aum,Basile:2023blg,Basile:2024dqq}
\begin{equation}
	c_i(\Lambda_s) \, \Lambda_s^2/M_P^2={\cal O}(1)\, ,
\end{equation}
such that 
\begin{equation}
	c_i(\Lambda_s) = N \, ,
\end{equation}
which provides the correct matching to the low energy effective action. 
This can be compared to the earlier obtained values of $g_0$ in specific compactifications, which shows that these coefficients are generically determined by the moduli-dependent number of particles $N$ in the definition of the species scale \cite{vandeHeisteeg:2022btw,Cribiori:2022nke,vandeHeisteeg:2023ubh,Cribiori:2023ffn,Cribiori:2023sch,vandeHeisteeg:2023dlw,Calderon-Infante:2023uhz,Castellano:2023aum,Basile:2023blg,Basile:2024dqq}.  Using the estimates of the species scale obtained in Ref.~\cite{vandeHeisteeg:2023dlw} this also leads to a matching of the unique effective action to string theory.

This implies, according to the swampland program, that at the boundary of the moduli space, namely in the large volume limit of the compact manifold or at weak coupling of the underlying string theory, the UV cut-off becomes parametrically smaller than $M_P$, which in turn is also relevant for the renormalization to smaller energies in the effective action context. 

We thus conclude that the Wilson coefficient of the second order terms in curvature $R^2$, $R_{\mu\nu} R^{\mu\nu}$, $R_{\mu\nu\rho\sigma} R^{\mu\nu\rho\sigma}$, $R\log\Box R$, $R_{\mu\nu}\log\Box R^{\mu\nu}$ and $R_{\mu\nu\rho\sigma}\log\Box R^{\mu\nu\rho\sigma}$ all have Wilson coefficients of order $N\sim M_P^2/\Lambda_s^2$.

Using these higher curvature operators and data from the Large Hadron Collider, one can set a limit on the number of species \cite{Alexeyev:2017scq} $N < 5 \times 10^{61}$. The strongest bound on $N$ comes directly from the Einstein-Hilbert term of the effective action.  From the non-observation  of quantum black holes at the Large Hadron Collider, one deduces $N<10^{32}$ \cite{Calmet:2008tn} which corresponds to an effective scale of quantum gravity in the TeV region.

{\it Conclusions}: In this work, we have considered the matching of the unique effective action to string theory. This matching enables us to calculate the values of the Wilson coefficients of the local part of this action at the Planck mass. The renormalization group equations for these coefficients enable us to run the values of these coefficients to a very low energy scale. This enables tests of string theory with low energy experiments or observations that are sensitive to gravitational physics. We thus show that actual predictions in string theory are possible, at least in the gravitational sector of the theory.

Additionally, we studied the situation where the theory contains a large number of species and described how the scale of quantum gravity depends on the species number, both in the effective field theory and string theory approaches, explaining how these arguments match. We discussed the value of the Wilson coefficients within this framework and its relation to the species scale $\Lambda_s=M_P/\sqrt{N}$. Moreover, we saw that all the Wilson coefficients of the effective field theory at second order in the curvature expansion are of order $N$.  Finally, we commented on current experimental limits on these operators.

\bigskip

{\it Acknowledgments:}
The work of X.C. is supported in part  by the Science and Technology Facilities Council (grants numbers ST/T006048/1 and ST/Y004418/1). E.K. thanks the Humboldt and Siemens foundations for support and the LMU group for their generous hospitality during the course of this work. The work of F.K. is supported by a postdoctoral fellowship of the Alexander von Humboldt foundation and a fellowship supplement of the Carl-Friedrich von Siemens foundation. The work of D.L. is supported by the Origins Excellence Cluster and by the German-Israel-Project (DIP) on Holography and the Swampland. 
\bigskip

{\it Data Availability Statement:}
This manuscript has no associated data. Data sharing not applicable to this article as no datasets were generated or analysed during the current study.

\bigskip

%\newpage


\begin{thebibliography}{10}
\bibitem{FN2} In string theory, the cosmological constant generically appears as a potential for matter scalar fields. From this perspective, we assume that we are sitting at a minimum of this potential with zero energy.
\bibitem{FN4} It is important to remember, that calculations in string theory are done mostly on shell, as the ones mentioned above. Therefore, effective Lagrangians related by field redefinitions are equivalent. In particular, by a field redefinition of the metric of the form $g_{\m\n}\to g_{\m\n}+aR_{\mu\nu}+bRg_{\m\n}$ one can remove the $R^2$ and $R_{\m\n}R^{\m\n}$ terms in the effective action. However this generates several other four-derivative terms of the other fields, that must be taken into account.
\end{thebibliography}
\end{document}